\documentclass[lettersize,journal]{IEEEtran}

\usepackage{amsmath,amsfonts}
\usepackage{amsthm}
\newtheorem{definition}{Definition}
\newtheorem{lemma}{Lemma}
\newtheorem{theorem}{Theorem}

\newcommand{\ft}{f}

\usepackage{algorithm}
\usepackage{array}
\usepackage[caption=false,font=normalsize,labelfont=sf,textfont=sf]{subfig}
\usepackage{textcomp}
\usepackage{stfloats}
\usepackage{url}
\usepackage{tabularx}
\usepackage{verbatim}
\usepackage{booktabs}
\usepackage{graphicx}
\usepackage{geometry}
\usepackage{svg}
\usepackage{color}
\usepackage{soul}
\usepackage{cuted}
\usepackage{cuted}
\usepackage[noend]{algpseudocode}
\algrenewcommand\algorithmiccomment[1]{\hfill\(\triangleright\)~#1}
\algrenewcommand\algorithmicrequire{\textbf{Input:}}
\algrenewcommand\algorithmicensure{\textbf{Output:}}
\usepackage{cite}
\usepackage{enumitem} 
\usepackage[colorlinks=true,citecolor=blue,linkcolor=black,urlcolor=blue]{hyperref} 
\begin{document}

\title{Update Strategy for Channel Knowledge Map in Complex Environments}

\author{Ting Wang, Chiya Zhang, Chang Liu, Zhuoyuan Hao, Rubing Han, Weizheng Zhang, and Chunlong He
\thanks{(Corresponding author: Chiya Zhang.)}
\thanks{This work was supported by the National Natural Science Foundation of China under Grant 62394294, and  Grant 62394290.}
\thanks{T. Wang, C. Zhang, C. Liu, R. Han, and W. Zhang are with the School of Information Science and Technology, Harbin Institute of Technology, Shenzhen, 518055, China. Z. Hao is with the School of Computer Science and Technology, Harbin Institute of Technology, Shenzhen, 518055, China.}
\thanks{C. He is with Guangdong Key Laboratory of Intelligent Information
Processing, Shenzhen University, Shenzhen, 518060, China.}}
\maketitle

\begin{abstract}
The Channel Knowledge Map (CKM) maps position information to channel state information, leveraging environmental knowledge to reduce signaling overhead in sixth-generation networks. However, constructing a reliable CKM demands substantial data and computation, and in dynamic environments, a pre-built CKM becomes outdated, degrading performance. Frequent retraining restores accuracy but incurs significant waste, creating a fundamental trade-off between CKM efficacy and update overhead.
To address this, we introduce a Map Efficacy Function (MEF) capturing both gradual aging and abrupt environmental transitions, and formulate the update scheduling problem as fractional programming. We develop two Dinkelbach-based algorithms: Delta-P guarantees global optimality, while Delta-L achieves near-optimal performance with near-linear complexity. For unpredictable environments, we derive a threshold-based policy: immediate updates are optimal when the environmental degradation rate exceeds the resource consumption acceleration; otherwise, delay is preferable. For predictable environments, long-term strategies strategically relax these myopic rules to maximize global performance. Across this regime, the policy reveals that stronger entry loss and faster decay favor immediate updates, while weaker entry loss and slower decay favor delayed updates.
\end{abstract}


\section{Introduction}
The evolution towards Sixth-Generation (6G) wireless networks is characterized by the deployment of extremely large-scale antenna arrays and the accommodation of a massive number of users\cite{10812728,10546919,8782879,electronics11081262}. This trend imposes unprecedented challenges on real-time channel state information (CSI) acquisition, where traditional channel estimation schemes suffer from prohibitive computational complexity and signaling overhead. To address this, the Channel Knowledge Map (CKM) has emerged as a promising paradigm for channel prediction. By constructing a digital twin of the wireless environment and leveraging offline training, a CKM can intelligently perceive the propagation environment and provide rapid online responses for channel parameter estimation, significantly reducing the overhead of real-time measurements\cite{arvai2020efficient}.
Substantial research has been dedicated to the construction of high-fidelity CKMs. Many existing work focus on utilizing detailed environmental information, such as the geometry and material properties of scatterers, to enhance prediction accuracy \cite{li2024enhancing}. These methods have proven effective in environments that are relatively static. 
\par
The efficacy of CKM is intrinsically linked to the strict correspondence between the stored map data and the actual physical propagation environment. Specifically, CKM's accuracy relies heavily on both the volume and timeliness of data, the latter of which is affected by environmental changes in the context of CKM application. Wireless environments are often dynamic, subject to changes such as the construction of new buildings, the movement of large vehicles, or even seasonal variations in foliage. Such temporal variations can render the existing CKM obsolete, leading to a severe degradation in prediction performance \cite{ting1}. This necessitates periodic CKM updates to maintain prediction fidelity, yet introduces new challenges in update management:
the prohibitive computational overhead arising from processing large-scale datasets for CKM update. 

Recent research has addressed \textit{how} to update CKMs when 
environmental changes occur. For instance, incremental learning techniques have been applied to adapt the CKM to new knowledge when environmental changes\cite{ting1}, such as the disappearance of old structures. However, conventional incremental learning may suffer from catastrophic forgetting where the model's memory of old knowledge is diluted as new data is continuously incorporated. To mitigate this, an unlearning-based fine-tuning mechanism was proposed in \cite{gong}, which demonstrated superior performance by selectively removing outdated information before adapting to new environmental features. While these methods effectively adapt CKMs to new environments, \textit{when} to trigger such updates to strike a balance between the improved performance and the cost overhead remains an open question.

Despite these advances in update mechanisms, a critical gap remains: when should updates be triggered? Unlike the ``how" question, which focuses on adaptation algorithms, the ``when" question involves strategic decision-making under resource constraints.
In typical urban scenarios, environmental changes occur at multiple timescales, while each CKM refresh requires much training data. For instance, retraining a neural network-based CKM for a single base station may involve collecting thousands of channel measurements and consuming minutes of computation on edge servers. This creates a non-trivial trade-off: \textbf{while frequent updates maintain accuracy, they incur substantial waste; conversely, delayed updates reduce waste but degrade prediction performance, directly impacting system throughput and reliability}.

The CKM update scheduling problem bears resemblance to 
information freshness management. In the domain of information-update systems, the Age of Information (AoI) and its variants have been widely adopted as metrics to quantify the freshness and value of information\cite{maatouk2023age,he2024age}. These metrics are instrumental in designing efficient scheduling policies for time-sensitive data. However, conventional AoI-based metrics primarily measure the time elapsed since the last update and do not inherently capture the performance degradation of a system caused by changes in the external world. They are not directly applicable to the CKM update problem, as they cannot characterize the impact of environmental dynamics on communication system performance.

To bridge this gap, we introduce a Map Efficacy Function (MEF) that evaluates CKM utility by incorporating environmental dynamics as a hidden variable. Based on this, we formulate the CKM update scheduling problem as an optimization problem aimed at maximizing the average long-term MEF of the CKM while minimizing the average resource consumption for updates. This formulation results in a fractional programming problem, for which we propose a two-layer iterative algorithm based on the Dinkelbach's method to find the optimal update strategy. Our work pioneers a systematic approach for deciding when to update a CKM, striking a balance between its operational performance and maintenance cost in complex wireless environments.

\subsection{Related Work}

\emph{1) Channel Knowledge Maps in Wireless Communications:}
The CKM serves as a comprehensive database linking geographic locations to multi-dimensional channel properties, enabling environment-aware networks for 6G \cite{zeng2021toward,Zeng2024Tutorial}. Existing research on CKM construction broadly divides into model-based methods leveraging physical propagation principles \cite{Xu2024HowMuchData} and data-driven techniques using machine learning \cite{Liu2023UAV}. However, most foundational work assumes quasi-static environments, limiting practical applicability. Recent efforts address CKM maintenance through statistical hypothesis testing to detect environmental changes \cite{Katagiri2022Dissertation} or by decomposing environments into quasi-static and dynamic components \cite{Wu2024EnvironmentAware}. 

\emph{2) Information Update System Update Strategy:}
The Age of Information (AoI) has emerged as a key metric for quantifying data timeliness \cite{yates2021age}, with extensive research on scheduling policies under various constraints \cite{sun2024aoi, tripathi2024whittle, kadota2019optimizing}. However, classic AoI is content-agnostic, prompting variants such as Age of Incorrect Information (AoII) and context-aware metrics like Value of Context-Aware Information (VoCAI) \cite{peng2024goal, li2024value}. These metrics fall short for CKM because: (1) CKM value stems from system-level prediction accuracy rather than source uncertainty; (2) our MEF captures performance degradation due to desynchronization with changing environments; (3) CKM updates involve substantial downtime during which utility is zero.

\emph{3) Dinkelbach Algorithm for Fractional Programming:}
Fractional programming, where objectives represent efficiency ratios, is common in wireless resource allocation \cite{crouzeix1991algorithms}. Dinkelbach's algorithm iteratively transforms such problems into tractable subtractive forms and has been applied to AoI optimization in recent work \cite{xiao2023adaptive}. We adopt this framework to formulate CKM update scheduling as maximizing the ratio of system utility to update cost.

Previous works \cite{bing, poker} explore optimal update timing within a single cycle, which corresponds to our short-term strategy in Section~IV. This paper systematically investigates \emph{when} to update CKM in dynamic environments, addressing the fundamental trade-off between CKM performance and retraining costs through fractional programming.

\subsection{Technical Challenges}
Specifically, this paper addresses the following technical challenges:
\begin{enumerate}[leftmargin=*]
\item The update problem lacks a model of environmental dynamics, in order to achieve a balance between the efficacy of CKM and the energy consumption of updates, We define a function MEF to quantify CKM utility over time in a complex environment and then formulate the CKM update scheduling problem as a long-term optimization.

\item  The formulated problem is a fractional scheduling with un-fixed variables. In order to solve it, we develop a Dinkelbach-Enabled Long-term Trajectory Algorithm with Pareto frontier(\textbf{Delta-P}), a two-parameter Dinkelbach algorithm with Pareto-frontier dynamic programming that propagates edge-additive triples \((F,G,C)\). We prove global optimality on the discrete candidate set, finite termination, and an \(O(\Delta)\) grid-approximation error to the continuous optimum. 

\item To reduce complexity, we propose a Dinkelbach-Enabled Long-term Trajectory Algorithm-Linearization(\textbf{Delta-L}), which employs Taylor linearization to reduce the inner problem to single-weight longest-path optimization on a DAG, achieving near-linear time in the number of feasible edges while preserving high solution quality.

\item  For unpredictable environments, we derive a threshold-based myopic policy using L'Hôpital's rule: immediate updates are optimal when environmental change rates exceed update cost acceleration, otherwise the waiting time could be calculated. We further characterize when long-term policies deviate from short-term ones due to strong entry loss and long subsequent segments.
\end{enumerate}
\section{System Model and Problem Formulation}
\label{sec:system-formulation}
We consider a single base station serving a fixed network task 
over a finite planning horizon in a controlled testbed or periodic scenario where environmental variations can be characterized.
A CKM provides channel prediction to support task execution. The CKM must be periodically updated through data recollection and model retraining. Crucially, each update incurs non-negligible duration and operational cost, during which the task is suspended and no value is accumulated. Following \cite{shisher2022does}, when each update incorporates 
all newly collected data, the prediction error decreases 
monotonically with update frequency. We adopt this full-update 
scheme to ensure that the CKM utility function $f(t)$ is 
monotonically decreasing in the age $t$ since the last update, 
which enables tractable analysis of the single-cycle optimization 
problem.
\par
Important symbols used are summarized in Table~\ref{tab:notation}.
\begin{table}[t]
  \caption{Notation and symbols}
  \label{tab:notation}
  \centering
  \begin{tabularx}{\linewidth}{@{}l l X@{}}
    \toprule
    \textbf{Symbol} & \textbf{Range/Unit} & \textbf{Meaning} \\
    \midrule
    $T_{\rm end}$ & time & Horizon end time \\
    $\mathcal{s}_j=[\tau_j,\tau_{j+1})$ & time & Update Segment $j$ \\
    $\ft(t)$ & $[0,1]$ & MEF \\
    $D(c)$ & $>0$ & Update resistance at time $c$ \\
    $C(c)$ & $\ge 0$ & Cost of completion at time $c$ \\
    $S=\{c_m\}$ & -- & Set of completion times, path of DAG \\
    $\mathcal{D}(S)$ & set & Downtime set \\
    $\mathcal{W}(S)$ & set & Working-time set\\
    $F(S)$ & -- & $\int_{\mathcal{W}(S)}\!\ft(t)\,dt$ \\
    $G(S)$ & time & $\mathrm{meas}(\mathcal{W}(S))$ \\
    $C_{\rm tot}(S)$ & -- & $\sum_m C(c_m)$ \\
    $H$ & time & $T_{\rm end}$ \\
    $J(S)$ & -- & Objective: average efficacy over working time minus average cost\\
    \bottomrule
  \end{tabularx}
\end{table}
\subsection{MEF: A New CKM Timeliness and Environment-State Metric}
The process of training and prediction within the CKM can be modeled as a virtual queue, with key events of ``Arrival" and ``Departure". In this model, the AoI of CKM represents the time elapsed since the most recent CSI was collected for the CKM currently in use. Specifically, AoI measures the ``freshness" of the CSI data, capturing the delay from the moment CSI is gathered to when it is processed through training, prediction, and subsequently applied in the communication network.

\par
Denote the arrival and departure time for packet $n$ as $T^A(n)$ and $T^D(n)$. $D$ is the processing time including $T_{train}$, $T_{wait2}$ and $T_{predict}$. Due to the different size of packets and different training process, $D$ may not be constant for the specific packet.
  \begin{figure*}[ht]
      \centering
      \includegraphics[width=0.8\textwidth]{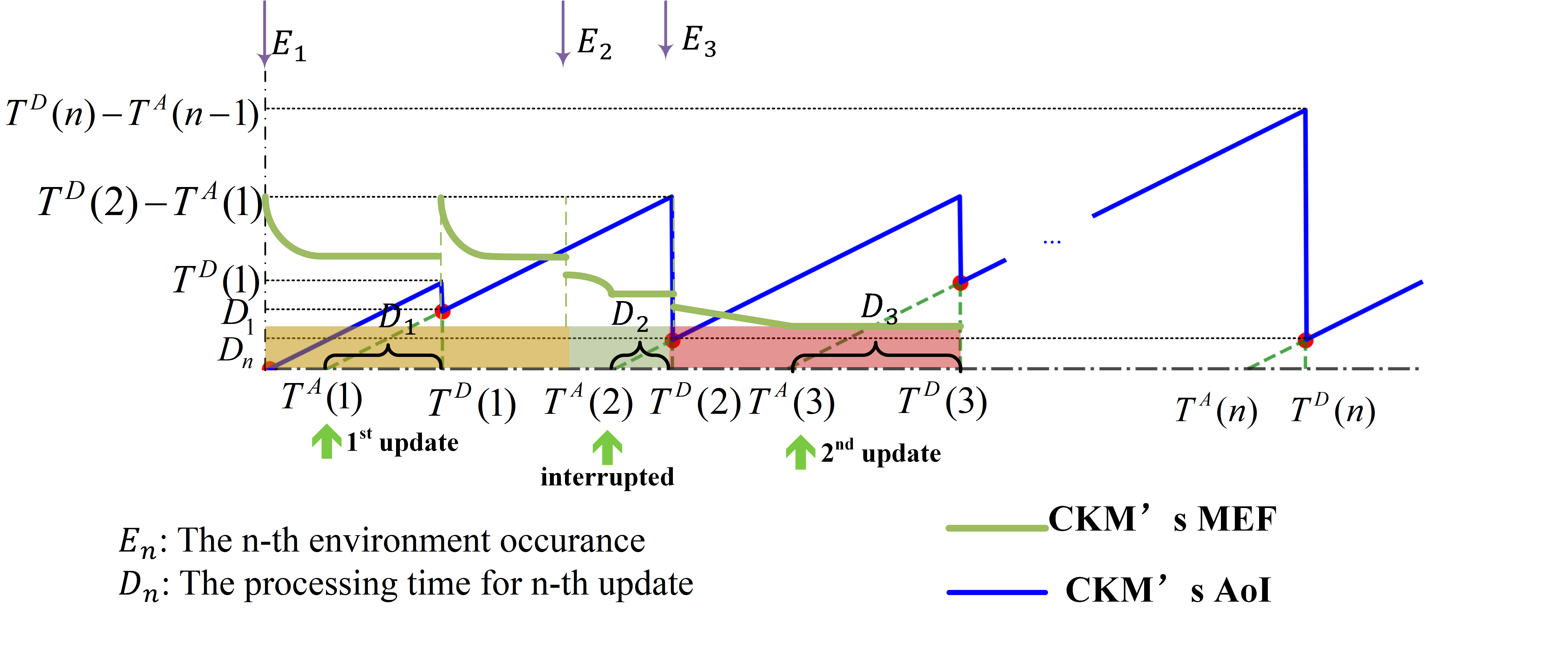}
      \caption{Evolution of AoI and MEF over update cycles.}
      \label{fig2}
    \end{figure*}
\par  
However, while AoI provides a linear measure of time elapsed since the last update, it may not be sufficient to capture the dynamic changes in the environment. We propose to use a function of information age, $f(t)\in [0,1]$, termed the map efficacy function (MEF) as our optimization metric, which can better reflect the value of CKM in dynamic environments. We define the $f(t)$ as the expected task performance ratio obtained when using the current, possibly outdated CKM relative to an ideal CKM hypothetically refreshed at time $t$.
\begin{equation}
  f(t) := \frac{\mathbb{E}[U(\pi(M_0)),\Theta_t]}{\mathbb{E}[U(\pi(M^*)),\Theta_t]}
\end{equation}
\par
Here, $M_{0}$ denotes the CKM currently in use, $M^{\star}$ is an ideal CKM that would be obtained if one could refresh instantaneously at time $t$. The mapping $\pi(\cdot)$ turns a CKM into the optimal network policy for the task, and $U(\pi,\Theta)$ is the corresponding task utility under environment state $\Theta_{t}$. The expectation $\mathbb{E}[\cdot]$ is taken over small-scale randomness conditional on $\Theta_{t}$. 
\par
$f(t)$ would decrease with the time since the last update finished. Fig. \ref{fig2} shows the AoI and MEF of CKM during several environment segments. AoI increases linearly with time and is reset to an initial value upon each updates. MEF decreases with time, displaying the fading value of CKM, and is also reset with the updates, during which it stays constant. If not reset, a sudden environment change will cause the MEF fading in another function with a time delay. Therefore, compared with AoI, MEF can better reflect not only the aging of the information but also the changes of the environment. 
\subsection{Environment Dynamics Model}
Since the communication environment could change due to sudden appearance or disappearance of base stations, weather, crowds and so on, the loss of CKM's efficacy in the environment does not always change in the same way. Define a correlation time for the communication environment, within which second-order channel statistics are essentially constant at the time scale of interest. Therefore, the horizon is partitioned into $M$ segments according to correlation time,
\begin{equation}
\mathcal{s}_j=[\tau_j,\tau_{j+1}),
0=\tau_1<\tau_2<\cdots<\tau_{M+1}=T_{\rm end},
\label{eq:segments}
\end{equation}
Segment boundaries model abrupt changes such as LOS$\leftrightarrow$NLOS transitions, weather onset and offset, vehicular or pedestrian surges.
\par
The formulation of MEF could be fitting by offline sampling, but for the theory analysis and simulation below, we adopt an analytical model.
We assume that inside segment $j$, the aging follows an exponential-to-floor law for two reasons: (1) the exponential decay $e^{-\lambda_j s}$ naturally models the gradual desynchronization between the CKM and the evolving environment, which is commonly observed in channel prediction systems where estimation error grows exponentially with information staleness \cite{shisher2022does}; (2) the floor $\eta_j$ captures the persistent environmental information (e.g., static building geometry) that remains valid despite temporal variations, preventing efficacy from vanishing entirely. In actual scenarios, other decay functions could be used without affecting the solution methodology.
\begin{align}
&E_j(s)\;=\;\eta_j+(1-\eta_j)\,e^{-\lambda_j s},\\
&\lambda_j=\frac{\ln 2}{T_{\mathrm{half},j}},\ \ \eta_j\in[0,1),
\label{eq:Ej}
\end{align}
where $T_{\mathrm{half},j}$ is the half-life and $\eta_j$ the persistent floor reflecting long-lived information. $E_j(s)$ describes the aging of a CKM generated at the beginning of $s_j$.
\par
To portray the impact of not updating the CKM in time when the environment updates, if a boundary $\tau_\ell$ is crossed without completing an update at the boundary, an instantaneous multiplicative boundary shock $s_\ell\in(0,1]$ is applied to the efficacy to account for sudden model mismatch at regime switches. Multiple consecutive unrefreshed boundaries compound multiplicatively. Formally, where $\ell$ indexes the segment boundaries defined in \eqref{eq:segments} with $\tau_\ell$ denoting the $\ell$-th boundary time, define the entry-loss process
\begin{equation}
L(t;S)\;=\;\prod_{\ell:\ \tau_\ell\in(t_{\rm last},\,t]}\, s_\ell,
\label{eq:entryloss}
\end{equation}
which resets to $1$ whenever a refresh completes exactly at a boundary.
\par
A refresh that completes at time $c$ occupies a downtime interval $[c-D(c),\,c)$ during which the CKM pipeline is not available for the task; consequently, no efficacy is produced. The downtime $D(c)>0$ and the cost $C(c)\ge 0$ are allowed to be segment-dependent, where $j(c)$ is the segment containing $c$. 
\par
A schedule is a finite, strictly increasing set of refresh completion times,
\begin{equation}
S=\{c_1<\cdots<c_K\}\subset[0,T_{\rm end}],
\label{eq:schedule}
\end{equation}
subject to non-overlapping downtimes:
\begin{equation}
c_1\ge D(c_1),
c_m-c_{m-1}\ge D(c_m),\ \ m=2,\dots,K.
\label{eq:nonoverlap}
\end{equation}
The task sees the new CKM only at $c_m$; the downtime that leads to $c_m$ is $[c_m-D(c_m),c_m)$. If $c_m=\tau_\ell$ (exactly at a boundary), the next segment starts with age $0$ and entry loss $1$; if the boundary is crossed without such a completion, the efficacy is multiplied by $s_\ell$.
\par
The downtime set induced by $S$ is
\begin{equation}
\mathcal{D}(S)\;=\;\bigcup_{m=1}^{K}[c_m-D(c_m),\,c_m),
\label{eq:downtime}
\end{equation}
and the working set is its complement in the horizon,
\begin{equation}
\mathcal{W}(S)\;=\;[0,T_{\rm end}]\setminus \mathcal{D}(S).
\label{eq:working}
\end{equation}
Combining the within-segment aging model $E_j(s)$ , the multiplicative entry-loss process $L(t;S)$, and the zero-efficacy constraint during downtime intervals $\mathcal{D}(S)$, we arrive at the operational form of the MEF $f(t)$ defined conceptually in \eqref{eq:ftilde}. Specifically, the age $s(t) = t - t_{\rm last}$ since the last completion, combined with the current segment index $j(t)$, determines the instantaneous efficacy:
\begin{equation}
f(t)\;=\;
\begin{cases}
0, & t\in \mathcal{D}(S),\\[4pt]
L(t;S)\,E_{j(t)}\!\big(s(t)\big), & t\in \mathcal{W}(S).
\end{cases}
\label{eq:ftilde}
\end{equation}
Equation \eqref{eq:ftilde} provides the operational form of the MEF defined conceptually in (1). While (1) characterizes efficacy abstractly as the expected performance ratio between the current CKM and an ideally refreshed one, \eqref{eq:ftilde} parameterizes this ratio through tractable components, enabling both theoretical analysis and numerical optimization of the update schedule $S$.

\textit{1) Assumptions.}
(i) $E_j(\cdot)$ is nonincreasing, continuous, and bounded in $[0,1]$; 
(ii) $\{S_\ell\}\subset(0,1]$ are known or pre-estimated (online learning methods would be discussed in another manuscript);
(iii) $D(\cdot),C(\cdot)$ are measurable and segment-dependent;
(iv) a feasible schedule exists. The parameters $\big\{(\eta_j,T_{{\rm half},j})\big\}$ can be estimated with some sampling data within segments; $D,C$ come from runtime and operational logs.

\textit{2) Normalization and units.}
All times are in consistent unit. $E_j(0)=1$ and $E_j(s)\downarrow \eta_j$ as $s\uparrow \infty$. The half-life $T_{{\rm half},j}$ is the time for the excess efficacy $(E_j(s)-\eta_j)$ to halve, it provides an interpretable decaying rhythm.

\medskip
\noindent\textit{Remarks.} 
(a) Setting $S_\ell\equiv 1$ removes boundary shocks, which means only the aging of CKM is considered. 
(b) In a stationary single-segment regime ($M=1$), the problem reduces to deciding when to update once. 
(c) Letting $\eta_j\to 0$ and $\lambda_j$ small yields $E_j(s)\approx 1-\lambda_j s$ for moderate ages, recovering AoI-like behavior after integration over working intervals. 
(d) Hard forbidden windows can be modeled by disallowing completions in specified intervals.

\subsection{Problem Formulation}
\label{subsec:problem}
We assess a schedule $S$ by a long-term utility-cost trade-off that averages efficacy while working and amortizes update cost through the whole time. The metric is designed based on the inspiration of the indicators used in industry to evaluate product performance.
\par
Let $\mathrm{meas}(\cdot)$ denote Lebesgue measure. The proposed objective is,
\begin{strip}
\begin{equation}
J(S)\;=\;
\underbrace{\frac{\displaystyle \int_{\mathcal{W}(S)}  f(t)\,dt}
{\displaystyle \mathrm{meas}(\bigl(\mathcal{W}(S)\bigr)}}_{\text{average normalized efficacy over working time}}
\;-\;
\underbrace{\frac{\displaystyle \sum_{m=1}^{K} C(c_m)}{T_{\rm end}}}_{\text{average update cost during the whole time}}
\label{eq:J}
\end{equation}
\end{strip}
The first term is the average efficacy of the CKM over its usable period, taking into account aging and losses caused by environment changes. The second  term is the update cost amortized over calendar time. This metric captures the long-term net value of the CKM, reflecting its balance between utility creation and resource consumption in actual operation, and is used to determine the most cost-effective update strategy.
\par
Given \eqref{eq:segments}--\eqref{eq:ftilde} and segment-dependent $D(\cdot),C(\cdot)$, the update scheduling problem is
\begin{equation}
\begin{aligned}
\mathbf{(P)}\qquad
&\underset{S=\{c_m\}\subset[0,T_{\rm end}]}{\text{maximize}} && J(S)\ \ \text{as in \eqref{eq:J}}\\
&\text{subject to} (\ref{eq:nonoverlap})
\end{aligned}
\label{prob:P}
\end{equation}

The solution of problem \textbf{(P)} yields the set of completion times that maximizes the long-term utility--cost tradeoff over $[0,T_{\rm end}]$.
\par
\section{Long-term Update Strategy for Predictable Environment}
We consider the offline, predictive setting in which segment-wise aging and entry-loss parameters, as well as segment-dependent overheads, are known a prior. We develop two algorithmic variants under the Dinkelbach framework: Delta-P (Dinkelbach with Pareto-frontier) for global optimality, and Delta-L (Dinkelbach with Taylor Linearization) for enhanced scalability. Both variants share the same outer fractional optimization structure but differ in their inner solvers.

Recalling the objective in \eqref{eq:J}, define
\begin{align}
&F(S)=\int_{\mathcal{W}(S)}  f(t)\,dt,\\
&G(S)=\mathrm{meas}(\bigl(\mathcal{W}(S)\bigr)),\\
&C_{\mathrm{tot}}(S)=\sum_{m=1}^{K} C(c_m),\\
&H=T_{\rm end}.
\end{align}
\par
The problem \textbf{(P)} can be reformulated as:
\begin{align}
    \mathbf{(P)}
    &\underset{S=\{c_m\}\subset[0,T_{\rm end}]}{\text{maximize}}  J(S)=\frac{F(S)H-C_{\mathrm{tot}}(S)G(S)}{G(S)H}\\
    &\text{subject to} (\ref{eq:nonoverlap}) \nonumber
\end{align}
The above problem is a fractional programming problem, which can be solved by the Dinkelbach algorithm\cite{Dinkelbach1967}.
\subsection{Delta-P: Dinkelbach-Enabled Long-term Trajectory Algorithm-Pareto Frontier}
We handle the two-ratio objective in \eqref{eq:J} with a two-parameter Dinkelbach scheme. For any feasible schedule $S$ with aggregated statistics $F(S),G(S),C_{\mathrm{tot}}(S),$ and $H$,
the goal is to maximize $J(S)$. The two-parameter Dinkelbach method maintains $(\lambda,\mu)$ as estimates of $(F/G,\ C_{\mathrm{tot}}/H)$ and, at iteration $k$, solves the parametric subproblem over the feasible set $\mathcal{X}$ induced by \eqref{eq:nonoverlap}:

\begin{equation}\label{eq:Phi}
\begin{aligned}
\Phi_{\lambda_k,\mu_k}(S) := &\; H\big(F(S)-\lambda_k G(S)\big) \\
&- G(S)\big(C_{\mathrm{tot}}(S)-\mu_k H\big) \longrightarrow \max_{S\in\mathcal{X}}.
\end{aligned}
\end{equation}

\begin{lemma}[Two-parameter Dinkelbach equivalence]
   The fractional program \textbf{(P)} can be solved by iteratively maximizing the parametric function $\Phi_{\lambda,\mu}(S)$ defined in \eqref{eq:Phi}, where $(\lambda,\mu)$ are updated as $\lambda \leftarrow F(S)/G(S)$ and $\mu \leftarrow C_{\mathrm{tot}}(S)/H$ until convergence.
   \end{lemma}
\begin{proof}
The classic parametric function of Dinkelbach is
\begin{align}
\nonumber
    \Phi_{\theta}(F,G,C_{tot})&=FH-C_{tot}G-\theta GH\\ \nonumber
    &=FH-C_{tot}G-(\lambda-\mu) GH\\ \nonumber
    &=H(F-\lambda G)-G(C_{\mathrm{tot}}-\mu H)\\
\end{align}
\end{proof}
While the optimization depends only on the composite parameter $\theta = \lambda - \mu$, we use a two-parameter formulation for its clearer physical interpretation. This approach decouples the marginal price of time $\lambda$ from that of update cost $\mu$, naturally reflecting the multi-objective trade-off between these distinct resources.
\par
 The multiplicative structure in $\Phi_{\lambda,\mu}$, namely the term $G(S)\,C_{\mathrm{tot}}(S)$, prevents a single-weight additive DP; we therefore rely on a Pareto-frontier mechanism that propagates path-wise statistics $(F,G,C_{\mathrm{tot}})$ and defers scalarization to selection. After solving the subproblem and obtaining $S_k$, the parameters are updated by
\begin{equation}
\lambda_{k+1}\;=\;\frac{F(S_k)}{G(S_k)},\qquad \mu_{k+1}\;=\;\frac{C_{\mathrm{tot}}(S_k)}{H},
\end{equation}
and the iteration stops when the Dinkelbach residual $|\Phi_{\lambda_k,\mu_k}(S_k)|$ falls below a tolerance. Under standard preconditions of positive denominator $G(S)>0$, nonempty feasibility, and exact solutions of the parametric subproblems, the sequence of objective values is monotonic and converges to the global optimum of $J(\cdot)$.
\par
The key algorithmic choice is to decouple the outer fractional update from the inner combinatorial structure. The outer loop performs a simple two-scalar update, while the inner loop solves a path problem on a directed acyclic graph (DAG) that encodes feasible update completions and working intervals. In each iteration, the inner solver builds a Pareto frontier of non-dominated $(F,G,C_{\mathrm{tot}})$ tuples at the sink and then selects the best tuple of $(\lambda_k,\mu_k)$ by maximizing the score
\begin{equation}
\text{score}(F,G,C\mid \lambda,\mu)\;=\;H\,F\;-\;G\,\big(C+(\lambda-\mu)H\big),
\end{equation}
which is algebraically identical to $\Phi_{\lambda,\mu}$ evaluated on $(F,G,C)$. This supports exact scalarization for any $(\lambda,\mu)$ without recomputing the entire DP, and also allows direct evaluation of $J=F/G-C/H$ on the frontier for diagnostics and early stopping.
\par
The initialization $(\lambda_0,\mu_0)$ can be set to zero, which is safe and simple. A stronger warm start uses the best static baseline to compute initial ratios. The stopping rule $|\Phi_{\lambda,\mu}(S)|<\texttt{tol}$ is robust; in practice we also check the absolute change of $J$ and guard against degenerate $G$ by enforcing a small lower bound on working time. Tie-breaking among schedules with identical scores is resolved by preferring smaller $C$ and then larger $G$.
\par
The pseudo-code of the whole algorithm is shown in Algorithm \ref{alg:gda-2param}. In each iteration, a Pareto-frontier DP produces the non-dominated triples $(F,G,C)$ for all feasible schedules on a DAG. The current iterate $(\lambda,\mu)$ linearize the fractional objective into $score(F,G,C\mid \lambda,\mu)$; then the algorithm select the frontier tuple that maximizes this score, update $\lambda \leftarrow F/G$ and $\mu \leftarrow C/H$ from the chosen schedule, and stop when the residual $\lvert\Phi\rvert$ falls below a tolerance. The frontier can be reused across iterations, while the scalarization changes only through $(\lambda,\mu)$.
\begin{algorithm}[t]
  \caption{Delta-P: Dinkelbach-Enabled Long-term Trajectory Algorithm-Pareto Frontier}
  \label{alg:gda-2param}
  \begin{algorithmic}[1]
  \Require Segment list $\mathcal{S}$; downtime $D$; per-update cost $C$; horizon $T_{\mathrm{end}}$; grid step $\Delta$; tolerance \texttt{tol}; max iterations \texttt{max\_iter}
  \Ensure Optimal completion times $S^\star$; objective $J^\star$
  \State $T \gets$ \textsc{BuildCandidateTimes}$(\mathcal{S}, T_{\mathrm{end}}, \Delta)$
  \State $(\mathcal{E}_{\mathrm{upd}},\mathcal{E}_{\mathrm{term}}) \gets$\textsc{PrecomputeEdges}$(T,\mathcal{S},T_{\mathrm{end}},D)$
  \State $\mathcal{P} \gets$ \textsc{ParetoFrontierDP}\!$(T,\mathcal{E}_{\mathrm{upd}},\mathcal{E}_{\mathrm{term}})$
  \State $\lambda\gets 0,\ \mu\gets 0,\ H\gets T_{\mathrm{end}}$
  \For{$i=1$ \textbf{to} \texttt{max\_iter}}
    \State $(S,F,G,C) \gets \arg\max \ \big[\,H\,F \;-\; G\,(C+(\lambda-\mu)H)\,\big]$
    \State $\Phi \gets H\,F \;-\; G\,(C+(\lambda-\mu)H)$
    \If{$|\Phi|<\texttt{tol}$} \State $S^\star\gets S$; $J^\star\gets F/G - C/H$; \textbf{break} \EndIf
    \State $\lambda \gets F/G$;\quad $\mu \gets C/H$
  \EndFor
  \State \Return $S^\star,\ J^\star$
  \end{algorithmic}
\end{algorithm}

\subsection{Parametric Subproblem and Pareto-Frontier DP Solver}
We encode feasible schedules as paths on a DAG whose vertex set $T$ collects all candidate times, including the segment boundaries $\{\tau_\ell\}$ and a uniform grid with step $\Delta$. There is a source at time $0$ and a sink at $T_{\mathrm{end}}$. An edge $(u\!\to\!v)$ is present if an update completion at $v$ is admissible: it must satisfy the non-overlap constraint $v-u\ge D(v)$ when the edge represents an update. We group edges into update edges and term edges. By retaining only edges that satisfy the constraints, feasibility is built into the graph; thus, any source-to-sink path corresponds to a feasible schedule with non-overlapping outages and a well-defined set of tasks. The goal is to transform the scheduling problem into a trajectory optimization problem on the graph, facilitating subsequent DP processing.
\par
For an \emph{update} edge $(u\!\to\!v)$,
\begin{flalign}
\label{eq:upd-inc}
&\Delta F_{u\to v}=\int_{u}^{\,v-D(v)}{f(t)}\,dt,\\
&\Delta G_{u\to v}=(v-D(v))-u,\\
&\Delta C_{u\to v}=C(v).
\end{flalign}
For a \emph{terminal} edge $(u\!\to\!T_{\mathrm{end}})$,
\begin{flalign}
\label{eq:term-inc}
&\Delta F_{u\to T_{\mathrm{end}}}=\!\!\int_{u}^{\,T_{\mathrm{end}}}\! L_{u\to T_{\mathrm{end}}}(t)\,E_{j(t)}\!(t-u)\,dt,\\
&\Delta G_{u\to T_{\mathrm{end}}}=T_{\mathrm{end}}-u,\\
&\Delta C_{u\to T_{\mathrm{end}}}=0.
\end{flalign}
Hence for any path $S$,
\begin{equation}
(F,G,C_{\mathrm{tot}})=\sum_{e\in S}(\Delta F_e,\Delta G_e,\Delta C_e),
\end{equation}
i.e., the three criteria remain additive across edges.
During downtime intervals, $ f(t)=0$ by definition and contributes neither to $\Delta F$ nor to $\Delta G$. These rules make $(\Delta F,\Delta G,\Delta C)$ Markovian given the current path semantics.
\par

Consequently, the inner problem is solved once as a
multi-criteria DP that maintains, at each vertex $v$, a set
$\mathcal{P}[v]$ of non-dominated tuples $(F,G,C)$ under the partial order.
\par
The DP proceeds in topological order. Initialize $\mathcal{P}[\text{src}]=\{(0,0,0)\}$ and $\mathcal{P}[v]=\emptyset$ for other vertices. For each edge $(u\!\to\!v)$ with increments $(\Delta F,\Delta G,\Delta C)$ and each tuple $(F_u,G_u,C_u)\in\mathcal{P}[u]$, form $(\hat F,\hat G,\hat C)=(F_u+\Delta F,\ G_u+\Delta G,\ C_u+\Delta C)$. Insert $(\hat F,\hat G,\hat C)$ into $\mathcal{P}[v]$ if it is not dominated; remove any tuples that it dominates. Keep backpointers for schedule recovery. At the sink, pass $\mathcal{P}[\text{sink}]$ to the outer loop and select the best tuple by maximizing $(F-\lambda G)-(C-\mu H)$ or, equivalently, by maximizing $J=F/G-C/H$ with $G>0$. To control the size of frontier sets, we optionally apply $\varepsilon$-dominance: $(F_1,G_1,C_1)$ $\varepsilon$-dominates $(F_2,G_2,C_2)$ if $F_1\ge (1-\varepsilon)F_2$, $G_1\le (1+\varepsilon)G_2$, and $C_1\le (1+\varepsilon)C_2$. This pruning yields a tunable complexity--accuracy trade-off and keeps memory usage predictable.
\begin{algorithm}[t]
  \caption{ParetoFrontierDP: Multi-criteria DP for $(F,G,C_{\mathrm{tot}})$}
  \label{alg:pareto-dp}
  \begin{algorithmic}[1]
  \Require DAG $(T,\mathcal{E})$; edge increments $(\Delta F,\Delta G,\Delta C)$
  \Ensure Frontier sets $\{\mathcal{P}[v]\}_{v\in T}$ with backpointers
  \State $\mathcal{P}[\text{src}] \gets \{(0,0,0)\}$;\quad $\mathcal{P}[v]\gets \emptyset$ for $v\neq \text{src}$
  \For{vertices $u$ in topological order}
    \For{each $(u\!\to\!v)\in \mathcal{E}$ with $(\Delta F,\Delta G,\Delta C)$}
      \For{each $(F_u,G_u,C_u)\in \mathcal{P}[u]$}
        \State $(\hat F,\hat G,\hat C)\gets (F_u+\Delta F,\ G_u+\Delta G,\ C_u+\Delta C)$
        \If{\textsc{NotDominated}$((\hat F,\hat G,\hat C),\ \mathcal{P}[v])$}
           \State Insert $(\hat F,\hat G,\hat C)$ with backpointer; prune dominated tuples in $\mathcal{P}[v]$
        \EndIf
      \EndFor
    \EndFor
    \State $\mathcal{P}[v]\gets$ \textsc{EpsilonPrune}$\big(\mathcal{P}[v],\ \varepsilon\big)$ 
  \EndFor
  \State \Return $\{\mathcal{P}[v]\}$
  \end{algorithmic}
\end{algorithm}
\subsection{Optimality}
The premise of the outer Dinkelbach optimality is that the inner algorithm has reached the optimal.
\subsubsection{Optimality of the Inner DP}
\paragraph{Sufficient conditions}
\begin{itemize}
  \item (C1) Additivity and nonnegativity: $(\Delta G_e,\Delta C_e,\Delta F_e)$ are additive and non-negative (cf. \eqref{eq:upd-inc}–\eqref{eq:term-inc}).\\
  \item (C2) Feasibility encoded and DAG: all constraints are encoded by feasible edges; the graph is acyclic and thus the complexity is controllable.\\
  \item (C3) Monotone scalarization: the final scoring $\Phi_{\lambda,\mu}$ is monotone nondecreasing in $F$ and nonincreasing in $G,C$, with $G>0$.
\end{itemize}
\paragraph{Roadmap}
Step 1: under (C1)–(C2), the DP computes the \emph{complete} Pareto frontier at the sink. 
Step 2: under (C3), any monotone scalarization $J$ and $\Phi_{\lambda,\mu}$ attains its global optimum on that frontier. 
Step 3: using Dinkelbach’s identity, the outer loop converges to the global optimum when the inner frontier is exact.

\paragraph{Setting}
On a DAG of discrete candidate times, each feasible path $S$ has an additive, nonnegative resource vector
\begin{align}
  &\mathbf u(S)=(G(S),\,C(S),\,F(S))=\sum_{e\in S}(\Delta G_e,\Delta C_e,\Delta F_e),\nonumber\\
  &(\Delta G_e,\Delta C_e,\Delta F_e)\in\mathbb{R}_+^3
\end{align}
\begin{definition}[Dominance, nondominance, Pareto frontier]
  Define the partial order
  \begin{equation}
\mathbf u_1\preceq \mathbf u_2\iff \big(G_1\le G_2,\ C_1\le C_2,\ F_1\ge F_2\big),
  \end{equation}
  with at least one strict inequality, we say $\mathbf u_1$ \emph{dominates} $\mathbf u_2$. 
  \par
  Let the terminal Pareto frontier be
\begin{equation}
\mathcal L_{\text{sink}}=\mathrm{ParetoMin}\big\{\mathbf u(S): S \text{ is a path to sink}\big\}
\end{equation}
\end{definition}

DP maintains at each node $i$ a Pareto label set $\mathcal L_i$. For any edge $e:i\to k$, extend $(g,c,f)\in\mathcal L_i$ by $(g+\Delta G_e,\,c+\Delta C_e,\,f+\Delta F_e)$ into $\mathcal L_k$, then prune dominated labels. Traverse once in topological order.

\paragraph{Objective and scalarizations}
For a single path vector $(g,c,f)$, define
\begin{align}
  \Psi(g,c,f)\ :=\ \frac{f}{g}-\frac{c}{H},\ \  (g>0,\ H>0).
  \end{align}
\begin{equation}
\Phi_{\lambda,\mu}(g,c,f)=H f-g\big(c+(\lambda-\mu)H\big).
\end{equation}

\begin{lemma}[Dominance preserved under extension]
\label{lem:ext}
If at node $i$, $\ell_1=(g_1,c_1,f_1)\preceq \ell_2=(g_2,c_2,f_2)$, then for any edge $e:i\to k$,
\begin{equation}
\ell_1+\Delta_e\ \preceq\ \ell_2+\Delta_e,\quad \Delta_e=(\Delta G_e,\Delta C_e,\Delta F_e).
\end{equation}
\end{lemma}
\begin{proof}
By (C1), additivity and componentwise nonnegativity imply
$g_1+\Delta G_e\le g_2+\Delta G_e$, $c_1+\Delta C_e\le c_2+\Delta C_e$, and $f_1+\Delta F_e\ge f_2+\Delta F_e$, with at least one strict. See standard properties of multiobjective shortest path label methods \cite{Ehrgott2005,Martins1984,Hansen1979}.
\end{proof}

\begin{lemma}[Safety of dominance pruning]
\label{lem:safe}
At the same node $i$, if $\ell_1\preceq \ell_2$, deleting $\ell_2$ discards no path that could become Pareto-optimal at $T$ or optimal under any (C3)-type scalarization.
\end{lemma}
\begin{proof}
For any suffix $P$ from $i$ to $sink$, apply Lemma \ref{lem:ext} edge by edge:
$\ell_1+\Delta(P)\preceq \ell_2+\Delta(P)$. Monotonicity in (C3) gives
$\Psi(\ell_1+\Delta(P))\ge \Psi(\ell_2+\Delta(P))$ and
$\Phi_{\lambda,\mu}(\ell_1+\Delta(P))\ge \Phi_{\lambda,\mu}(\ell_2+\Delta(P))$.
Thus dominated labels can be safely removed.
\end{proof}

\begin{lemma}[Completeness invariant]
\label{lem:inv}
In topological order, $\mathcal L_i$ equals the Pareto frontier of all feasible labels reaching $i$.
\end{lemma}
\begin{proof}
At the source, $\mathcal L_{\mathrm{src}}=\{(0,0,0)\}$. Assume it holds for all predecessors of $i$. Extending their frontiers along feasible edges covers all labels that reach $i$; pruning via Lemma \ref{lem:safe} yields the Pareto frontier at $i$.
\end{proof}

\begin{theorem}[Completeness of the terminal frontier]
\label{thm:complete}
Under (C1)–(C2) and a finite DAG, upon termination $\mathcal L_T$ is exactly the Pareto frontier of all feasible paths to $T$.
\end{theorem}
\begin{proof}
Apply Lemma \ref{lem:inv} at $i=sink$. Finiteness and acyclicity ensure termination in one topological pass.
\end{proof}

\begin{theorem}[Global optimality attained on the frontier]
\label{thm:select}
Under (C3) with $G>0$,
\begin{equation}
\max_{S} J(S)=\max_{\mathbf u\in \mathcal L_T}\Psi(\mathbf u),\ \
\max_{S}\Phi_{\lambda,\mu}(S)=\max_{\mathbf u\in \mathcal L_T}\Phi_{\lambda,\mu}(\mathbf u).
\end{equation}
\end{theorem}
\begin{proof}
Monotonicity: if $\mathbf u_1\preceq \mathbf u_2$ and $g_1,g_2>0$, then
\begin{align}
&\frac{f_1}{g_1}-\frac{f_2}{g_2}
=\frac{f_2(g_2-g_1)+(f_1-f_2)g_2}{g_1g_2}\ \ge\ 0,\\
&-\frac{c_1}{H}\ \ge\ -\frac{c_2}{H},
\end{align}
hence $J(\mathbf u_1)\ge J(\mathbf u_2)$, and similarly for $\Phi_{\lambda,\mu}$. Any non-Pareto label is dominated by a Pareto label with no worse score; therefore the maxima are reached on $\mathcal L_T$.
\end{proof}
\subsubsection{Optimality of the Dinkelbach Algorithm}
$H>0$ and $G(S)>0$ hold for all feasible paths $S$. According to the definition of $J(S)$ and $\Phi_{\lambda,\mu}(S)$,with $\theta:=\lambda-\mu$,
\begin{equation}\label{eq:phi-eq}
\Phi_{\lambda,\mu}(S)=G(S)H\big(J(S)-\theta\big).
\end{equation}

\begin{lemma}[Sign equivalence]\label{lem:sign}
For any $(\lambda,\mu)\in\mathbb{R}^2$ and any feasible $S$,
$\ \Phi_{\lambda,\mu}(S)\ge 0\iff J(S)\ge \lambda-\mu$.
\end{lemma}
\begin{proof}
Immediate from \eqref{eq:phi-eq} and $G(S)H>0$.
\end{proof}

\begin{lemma}[Residual nonnegativity and exact stopping]\label{lem:residual}
Given $(\lambda_k,\mu_k)$, let
\begin{equation}
S_k\in\arg\max_{S}\Phi_{\lambda_k,\mu_k}(S),\qquad
r_k:=\Phi_{\lambda_k,\mu_k}(S_k),
\end{equation}
and update $\lambda_{k+1}=F(S_k)/G(S_k)$, $\mu_{k+1}=C(S_k)/H$ (equivalently, $\theta_{k+1}=J(S_k)$).
Then $r_k\ge 0$, and $r_k=0$ if and only if $S_k$ is a global maximizer of $J$.
\end{lemma}
\begin{proof}
By the update at iteration $k-1$, $\theta_k=\lambda_k-\mu_k=J(S_{k-1})$. Hence by \eqref{eq:phi-eq},
$\Phi_{\lambda_k,\mu_k}(S_{k-1})=0$, so $r_k=\max_S\Phi_{\lambda_k,\mu_k}(S)\ge 0$.
Moreover, $r_k=0$ iff $\Phi_{\lambda_k,\mu_k}(S)\le 0$ for all $S$ and equality holds at $S_k$.
By Lemma~\ref{lem:sign}, this is equivalent to $J(S)\le \theta_k$ for all $S$ and $J(S_k)=\theta_k$,
i.e., $S_k$ attains the global maximum of $J$.
\end{proof}

\begin{lemma}[Monotone ascent]\label{lem:ascent}
The sequence $\{J(S_k)\}$ is nondecreasing, and strictly increasing whenever $r_k>0$.
\end{lemma}
\begin{proof}
By \eqref{eq:phi-eq}, $r_k=G(S_k)H\big(J(S_k)-\theta_k\big)\ge 0$, so $J(S_k)\ge \theta_k=J(S_{k-1})$.
If $r_k>0$, then $J(S_k)>\theta_k$, hence strict ascent.
\end{proof}

\begin{lemma}[Finite termination under discreteness]\label{lem:finite}
If the feasible set is finite, the iteration terminates in finitely many steps with $r_k=0$ and $J(S_k)=\max_S J(S)$.
\end{lemma}
\begin{proof}
By Lemma~\ref{lem:ascent}, $\{J(S_k)\}$ is a nondecreasing sequence taking values in a finite set, hence becomes stationary in finitely many steps; by Lemma~\ref{lem:residual}, stationarity implies $r_k=0$ and global optimality.
\end{proof}

\begin{theorem}[Global optimality of the two-parameter Dinkelbach scheme]\label{thm:dinkelbach}
Let $J^\star=\max_S J(S)$. With exact inner maximization in each iteration, the sequence $\{J(S_k)\}$ is nondecreasing and converges to $J^\star$, with $r_k\to 0$. If, in addition, $J$ takes only finitely many values on the feasible set, the algorithm terminates finitely at a global maximizer of $J$.
\end{theorem}
\begin{proof}
By Lemma~\ref{lem:ascent}, $\{J(S_k)\}$ is nondecreasing and bounded above by $J^\star$, hence convergent to some $\bar J\le J^\star$. If $\bar J<J^\star$, exact inner maximization together with \eqref{eq:phi-eq} would permit strictly positive residuals infinitely often, contradicting convergence; thus $\bar J=J^\star$ and $r_k\to 0$. The finite termination case follows from Lemma~\ref{lem:finite}.
\end{proof}

\subsection{Convergence}
The superlinear convergence of Dinkelbach's Algorithm stems from the fact that it is
equivalent to a Newton’s method applied to a convex function derived from the fractional objective\cite{crouzeix1991algorithms}.
\par
\begin{theorem}[Inner DP Convergence]\label{thm:dag_converge}
On a finite DAG with nonnegative edge weights $(\Delta F,\Delta G,\Delta C)$, 
the Algorithm~2 computes the exact Pareto frontier 
$\mathcal{P}[\mathrm{sink}]$ in one topological pass and terminates finitely.
\end{theorem}

\begin{proof}
Any path uses at most $|V|-1$ edges; 
each $(F,G,C)$ is a finite sum of nonnegative increments, so the label space is finite.
\end{proof}

\subsection{Complexity}
The time complexity of Delta-P is proportional to the number of edges times the average frontier size. On a grid with $|T|$ vertices and $|\mathcal{E}|$ edges, the worst-case cost is $O(|\mathcal{E}|\,\Gamma)$ where $\Gamma$ bounds the per-vertex frontier cardinality after pruning; in practice, $\Gamma$ remains moderate under realistic parameters and coarse-to-medium grids. The method scales linearly with the number of Dinkelbach iterations, which is typically small. 

\subsection{Delta-L: Dinkelbach Enabled Long Trajectory Algorithm-Linearization}
When $\Gamma$ grows (fine grids or many constraints), Delta-P becomes the bottleneck. To reduce complexity, we propose Delta-L, which adopts Taylor expansion for the multiplicative term in Eq.~\ref{eq:Phi},
which contains the non-additive product $G\,C_{\mathrm{tot}}$ at the path level.
\par
At inner iteration $k$, given the current schedule statistics $(F^{(k)},G^{(k)},C^{(k)})$, linearize the product $f(G,C)=G\,C$ at the current iterate $(G^{(k)},C^{(k)})$ using the first-order Taylor expansion:
\begin{small}
\begin{align}
&G\,C \;\approx\; f(G^{(k)},C^{(k)}) + \frac{\partial f}{\partial G}\Big|_{(k)}(G-G^{(k)}) + \frac{\partial f}{\partial C}\Big|_{(k)}(C-C^{(k)})\\\nonumber
&\space\space\space\space\space\space\space\space= C^{(k)}G + G^{(k)}C - G^{(k)}C^{(k)}.
\end{align}
\end{small}
Dropping the constant, the path score becomes edge-additive with per-edge weight
\begin{equation}
w_k(e)\;=\;H\,\Delta F_e\ -\ \big(C^{(k)}+(\lambda-\mu)H\big)\,\Delta G_e\ -\ G^{(k)}\,\Delta C_e.
\end{equation}
Thus one Taylor Linearization (TL) step in Delta-L solves a single-weight longest-path DP on the DAG. Overall time complexity is shown in Table \ref{tab:oracle-compare}.
\begin{table*}[t]
  \caption{Complexity and optimality: Delta-P vs. Delta-L}
  \label{tab:oracle-compare}
  \centering
  \begin{tabular}{@{}lccccc@{}}
    \toprule
    Method & Inner time & Total time & Space & Outer reuse & Optimality \\
    \midrule
    Delta-P & $O(|\mathcal{E}|\,\Gamma)$ & $O(N_{\text{outer}}+|\mathcal{E}|\,\Gamma)$ & $O(|T|\,\Gamma)$ & reuse frontier; select in $O(\Gamma)$ & global \\
    Delta-L & $O(|\mathcal{E}|)$ & $O(N_{\text{outer}}\,N_{\text{TL}}\,|\mathcal{E}|)$ & $O(|T|)$ & reweight each TL step & approximate \\
    \bottomrule
  \end{tabular}
\end{table*}
\par
\textbf{\textit{example}}
Let
\begin{small}
  \begin{equation}
T=\{t_0=0<t_1=2<t_2=5<t_3=9<t_4=12=T_{\rm end}\},
\end{equation}
\end{small}
and include an edge $u\!\to\!v$ iff $v>u$ and $v-u\ge D(v)=2$. Precompute edge increments $(\Delta F,\Delta G,\Delta C)$ as in \eqref{eq:upd-inc}--\eqref{eq:term-inc}. 
Then feasible edges include
\begin{align}
&\mathcal{E}(t_0)=\{t_0\!\to\!t_1,\ t_0\!\to\!t_2,\ t_0\!\to\!t_3,\ t_0\!\to\!t_4\},\\
&\mathcal{E}(t_1)=\{t_1\!\to\!t_2,\ t_1\!\to\!t_3,\ t_1\!\to\!t_4\},\\
&\mathcal{E}(t_2)=\{t_2\!\to\!t_3,\ t_2\!\to\!t_4\},\\
&\mathcal{E}(t_3)=\{t_3\!\to\!t_4\},
\end{align}
and every $u\in T$ may also have a direct terminal edge $u\!\to\!T_{\rm end}$.
\par
\noindent The DP traverses vertices in the topological order
\begin{equation}
t_0\;\rightarrow\;t_1\;\rightarrow\;t_2\;\rightarrow\;t_3\;\rightarrow\;t_4(=T_{\rm end}),
\end{equation}
Initialize the frontier at the source $P[t_0]=\{(0,0,0)\}$. For each vertex $u$ in this order and each outgoing edge $(u\!\to\!v)$ with increments $(\Delta F,\Delta G,\Delta C)$, propagate candidates
\begin{equation}
 (\hat F,\hat G,\hat C)=(F+\Delta F,\ G+\Delta G,\ C+\Delta C)
\end{equation}
from $P[u]$ to $P[v]$, inserting only non-dominated tuples. After processing $t_4$, select from $P[t_4]$ the tuple that maximizes the current scalarization, and recover the corresponding path by backpointers.
\par
One TL iteration proceeds as:
\begin{enumerate}[leftmargin=1.5em,itemsep=2pt]
\item Given $(G^{(k)},C^{(k)})$, assign each edge $e$ the weight
$w_k(e)=H\,\Delta F_e-\big(C^{(k)}+(\lambda-\mu)H\big)\Delta G_e-G^{(k)}\Delta C_e$.
\item Run a single-weight longest-path DP in the topological order $t_0\!\to t_1\!\to t_2\!\to t_3\!\to t_4$:
\begin{equation}
\mathrm{dp}[t_0]=0,\qquad \mathrm{dp}[v]=\max_{u\to v}\ \big(\mathrm{dp}[u]+w_k(u\!\to\!v)\big),
\end{equation}
keeping backpointers to recover the path $S^{(k+1)}$.
\item Update $(F^{(k+1)},G^{(k+1)},C^{(k+1)})$ by summing the edge increments on $S^{(k+1)}$. 
\item Stop the inner loop if $|\Phi_{\lambda,\mu}(S^{(k+1)})-\Phi_{\lambda,\mu}(S^{(k)})|<\texttt{tol}_{\rm in}$ or $S^{(k+1)}=S^{(k)}$; otherwise set $k\!\leftarrow\!k{+}1$ and repeat.
\end{enumerate}
In the outer two-parameter Dinkelbach loop, update $\lambda\!\leftarrow\!F/G$ and $\mu\!\leftarrow\!C/H$, and stop when $|\Phi_{\lambda,\mu}(S)|<\texttt{tol}$. Compared with the Delta-P, the Delta-L surrogates the frontier maintenance by a few $O(|\mathcal{E}|)$ longest-path solves, trading exactness for speed in regimes where $\Gamma$ is large.

\section{Short-term Update Strategy for Unpredictable Environments}
\label{sec:short-term}
In non-predictive settings, where future environmental changes are unknown, the update strategy becomes short-term. The decision to update is based solely on the currently observed efficacy decay and known overheads. Let $D>0$ be the update downtime and $C\ge 0$ its cost. The problem reduces to finding the optimal time $t$ for a single update to maximize the net utility, formulated as:
\begin{equation}
  \max \limits_{t} g(t)=\frac{\int_0^tf(x)dx}{t}-\frac{C}{t+D}
\end{equation}
\par 
Optimizing $g(t)$ yields the optimal update moment based on the current observed values. Since we don't know how long each environment exists, just one update is decided for once optimization.
\par
According to the properties of $f(t)$, the results differ in different situations, but in total it could be concluded that the optimal update time is a threshold structure: when $\frac{C}{D^2} \leq -\frac{f'(0)}{2}$, the optimal update time is $t=0$, otherwise, the optimal update time is the first $t^*$ when it satisfies $g'(t^*)=\frac{tf(t)-\int_0^tf(x)dx}{t^2}+\frac{C}{(t+D)^2}=0$. 
\par
The derivative of $g(t)$ is a sum of two parts, one of which is evidently positive or zero and the other is evidently negative or zero. If the sum of the two terms is zero, the $t$ is the optimal update time. If $f(t)$ is constant, $g(t)$ is increasing over time, which means an update is always unnecessary. If $C=0$, that is, the update is free, $g(t)$ is decreasing over time and the optimal update time is $t=0$, which means an update could be performed ever since the environment changes. In other normal cases, $g'(t)$ is a sum of one positive part and one negative part. 
\par
Let $h(t)=tf(t)-\int_0^tf(x)dx$, from the above we know that $h(t)< 0$ and $h'(t) < 0$. let $p(t)=\frac{C}{(t+D)^2}$, it's evident that $p(t)$ is a decreasing function which declines from a value of $\frac{C}{D^2}$ and asymptotically approaches zero. To discuss the positive and negative shapes of $g'(t)$, we only need to consider the rates of change of $\frac{h(t)}{t^2}$ and $p(t)$. 
\par
First, let's study the value of $\frac{h(t)}{t^2}$ at t=0. When $t$ is close to 0, $h(t)$ and $t^2$ are also close to 0, so L'Hôpital's Rule is applied:
\begin{align}
    \lim_{t\to0}H(t)=\lim_{t\to0}\frac{h(t)}{t^2}&=\lim_{t\to0}\frac{tf(t)-\int_0^tf(x)dx}{t^2}\\\nonumber
    &=\lim_{t\to0}\frac{tf'(t)}{2t}\\\nonumber
    &=\lim_{t\to0}\frac{f'(t)}{2}\nonumber
\end{align}
The positive part is always a decreasing function from a positive value to close to zero, and the negative part could be a decreasing function from a negative value or zero to $-\infty$ or a negative value, or a increasing function from a negative value or $-\infty$ to a negative value or zero.
\par
\begin{figure*}[t]
  \centering
  \includegraphics[width=\linewidth]{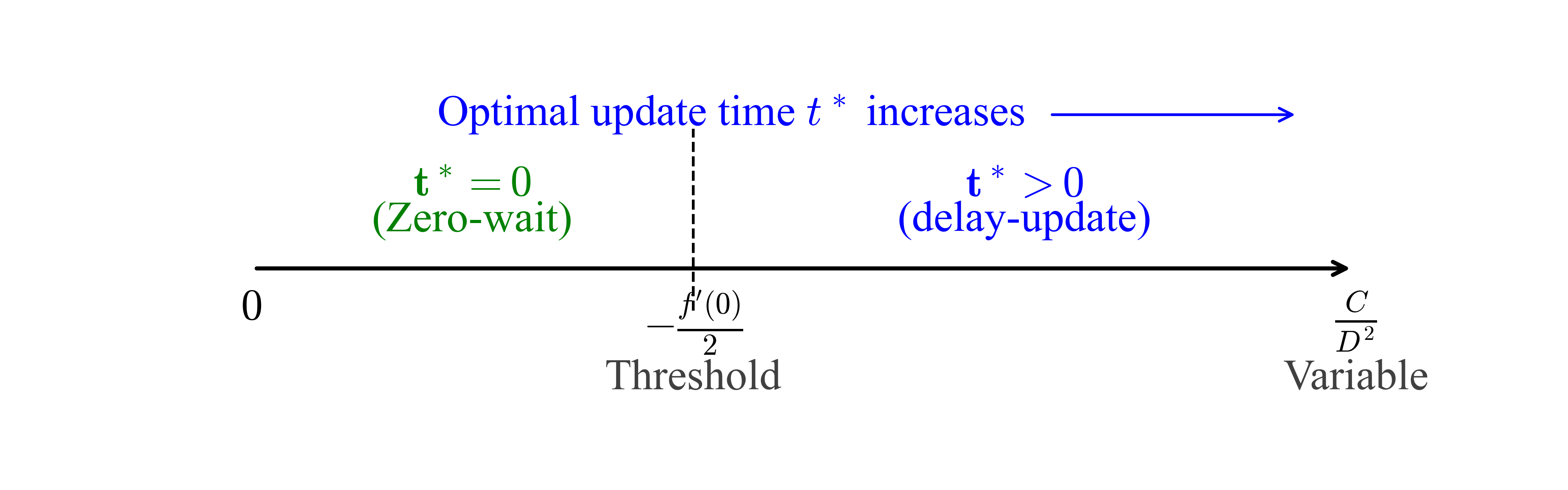}
  \caption{Illustration of the threshold-based update strategy. The horizontal axis represents the acceleration of \textbf{computational resource consumption} ($\frac{C}{D^2}$), while the vertical dashed line denotes the threshold determined by the \textbf{environmental change rate} ($-\frac{f^{\prime}(0)}{2}$). The zero-wait policy is optimal only when the acceleration of computational resource consumption is lower than the rate of environmental degradation.}
  \label{fig4}
  \end{figure*}

Based on the above analysis, we can draw the final conclusion, as shown in Fig.\ref{fig4}.
\par
We can interpret the threshold structure of the optimal update strategy from a practical perspective. The ratio $\frac{C}{D^2}$ can be understood as the rate of change in computational resource consumption during updates, which reflects how quickly the system's computing power is consumed relative to the update delay. This ratio serves as a critical threshold that determines whether immediate updates are optimal.
\par
When the environment changes rapidly (characterized by a large negative value of $\frac{f'(0)}{2}$), the information value deteriorates quickly over time, making it more valuable to maintain fresh information despite the computational costs. In this scenario, the optimal strategy is to update immediately ($t^*$ = 0), as the benefit of maintaining information freshness outweighs the computational overhead. This is intuitively reasonable because in rapidly changing environments, the cost of using stale information for decision-making would be significantly higher than the computational cost of frequent updates.
Conversely, when the environment changes slowly, that is, when the value of $-\frac{f'(0)}{2}$ is small, the information remains valuable for a longer period, and the system can afford to wait longer between updates to minimize computational resource consumption. The threshold structure thus provides a way to balance the trade-off between information freshness and computational efficiency based on the relative rates of environmental change and resource consumption.
\section{Simulation}
\subsection{Simulation Setup}
We consider a horizon of $T_{\rm end}=300$ minutes with $M=6$ segments per case. 
We adopt a hierarchical grid: a coarse mesh with $\Delta_{\rm coarse}=5$ minutes across $[0,T_{\rm end}]$, augmented by a fine mesh with $\Delta_{\rm fine}=0.25$ minutes within $\pm 12$ minutes of every boundary $\{0,\tau_j,T_{\rm end}\}$, while always including the boundaries themselves. After a first pass, we perform a single local refine-and-resolve step by injecting extra candidates within $\pm 6$ minutes around the obtained completion times using a $0.2$-minute step and rerunning the solver once; this acts as an inexpensive, one-shot adaptive refinement that aligns the continuous optimum to the discrete grid.
\par
To further control complexity without sacrificing optimality, we prune very long edges by enforcing $t_k-t_i\le 90$ minutes. All per-edge increments $(\Delta F,\Delta G,\Delta C)$ are computed exactly once by closed-form, boundary-aware integration and reused across inner and outer iterations, eliminating repeated numerical integration. 
\par
To verify the characteristics of our algorithm in different environments, We draw $N$ i.i.d. cases for each of three environment types summarized in Table~\ref{tab:env-types}; in all types the per-update downtime $D$ and cost $C$ are segment-dependent.

\begin{table*}[t]
  \caption{Environment types and sampling bands used in simulation}
  \label{tab:env-types}
  \centering
  \begin{tabular}{@{}lcccc@{}}
    \toprule
    Type & $S_{\rm entry}$ & $T_{\!1/2}$ (min) & $\eta$ & $(D,\ C)$ (typical) \\
    \midrule
    A: small-entry, fast & $[0.2,0.5]$ & $[2.5,5]$ & $[0,0.08]$ & $D\in\{0.8,1.0,1.2,1.5\}$,\ $C\in[0.05,0.2]$ \\
    B: large-entry, slow & $[0.9,1.0]$ & $[50,80]$ & $[0.2,0.35]$ & $D\in[2.0,3.5]$,\ $C\in[0.8,1.8]$ \\
    C: mixed off-diagonals & $\{[0.6,0.8]\ \text{or}\ [0.9,1.0]\}$ & $\{[50,80]\ \text{or}\ [8,18]\}$ & $[0.05,0.30]$ & $D\in[2.0,3.5]$,\ $C\in[0.8,1.8]$ \\
    \bottomrule
  \end{tabular}
\end{table*}
\par
For the numerical solver, we set the outer Dinkelbach tolerance to $10^{-6}$ with a maximum of 60 iterations, and the inner Taylor-linearization (TL) solver tolerance to $10^{-6}$ with up to 25 iterations per outer step. The grid is initialized with $\Delta_{\rm coarse} = 3$ minutes, refined locally around boundaries and obtained completion times with $\Delta_{\rm fine} = 0.2$ minutes, ensuring that all segment boundaries $\{\tau_j\}$ are included as candidate times. To reduce edge count, we prune edges spanning more than 90 minutes. All edge increments $(\Delta F, \Delta G, \Delta C)$ are precomputed via closed-form integration of \eqref{eq:upd-inc}--\eqref{eq:term-inc} and reused across iterations.

\subsection{Simulation Results}
We first compare our near-optimal strategy with other three policies:(1)\textbf{Zero-wait Update}: update once the environment get into a new segment. (2)\textbf{Fixed-10m}: update every 10mins since the initial start. (3)\textbf{Fixed-25m}: update every 25mins since the initial start.
\par
To diagnose local one-cycle incentives, for each segment $j$ we compute the initial decay slope and its short-term threshold
\begin{equation}
T_j = -\frac{f_j'(0)}{2}=\frac{(1-\eta_j)\lambda_j}{2},
\end{equation}
which is displayed together with $(D_j,C_j)$ in the trajectory figures. The short-term sufficient rule “zero-wait if $R_j=C_j/D_j^2\le T_j$” serves as a reference line.

\begin{figure}[h]
    \centering
    \includegraphics[width=0.5\textwidth]{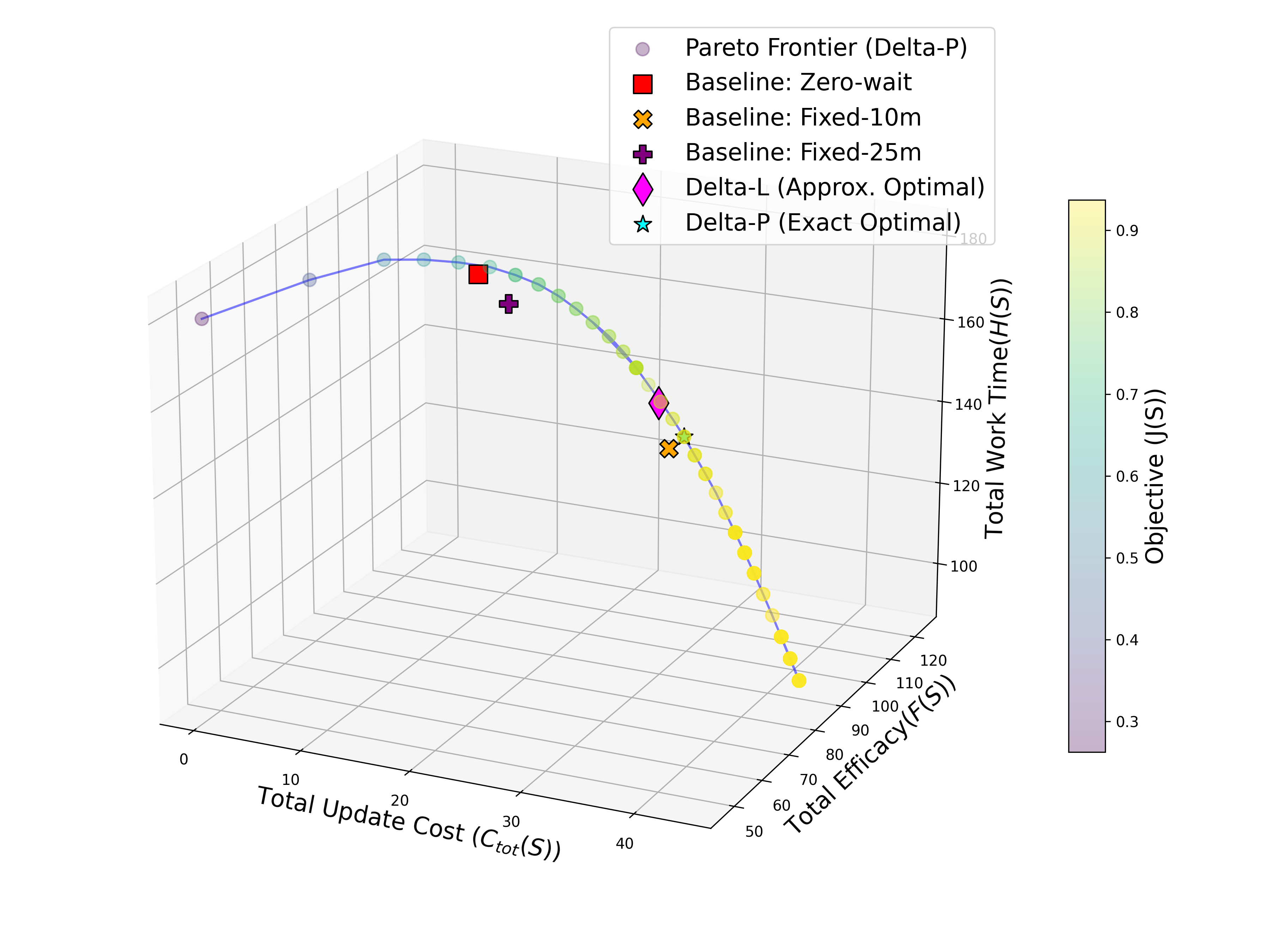} 
    \caption{Pareto frontier of working time $G$, efficacy $F$, and update cost $C_{tot}$.}
    \label{fig:pareto_frontier_3d}
\end{figure}
\par
Fig. \ref{fig:pareto_frontier_3d} illustrates the performance of various CKM update strategies against the Pareto frontier, which represents the theoretical performance limit defining the optimal trade-off between total update cost, efficacy score, and working time. The results validate our proposed Delta-P (cyan star), precisely achieves this optimal frontier, confirming its theoretical optimality. Crucially, the low-complexity Delta-L algorithm (magenta diamond) yields a solution remarkably close to the frontier, demonstrating its near-optimal performance. In contrast, all baseline strategies are shown to be sub-optimal, as their performance points lie significantly inside this boundary.

\begin{table*}[htbp]
    \centering
    \caption{Performance of Delta-P/L and Baseline Strategies}
    \label{tab:performance_comparison}
    \begin{tabular}{lrrrrr}
        \toprule
        \textbf{Strategy} & Efficacy & Work Time & Update Cost &  \textbf{Objective}  & Computation Time\\
        \midrule
        \textbf{Delta-P}  & 151 & 167&31.5&\textbf{0.7674}&63.62s\\
        \textbf{Delta-L} & 153 & 173 & 28.5 &  \textbf{0.7628} & 0.45s\\
        Zero-wait        & 130 & 215 &  7.5 &  0.5731 & 0.00s\\
        Fixed-10m        & 143 & 164 & 33 &  0.7308 & 0.00s\\
        Fixed-25m        & 145 & 203 & 13.5 &  0.6563 &0.00s\\
        \bottomrule
    \end{tabular}
\end{table*}



\begin{figure*}
    \centering
    \includegraphics[width=\linewidth]{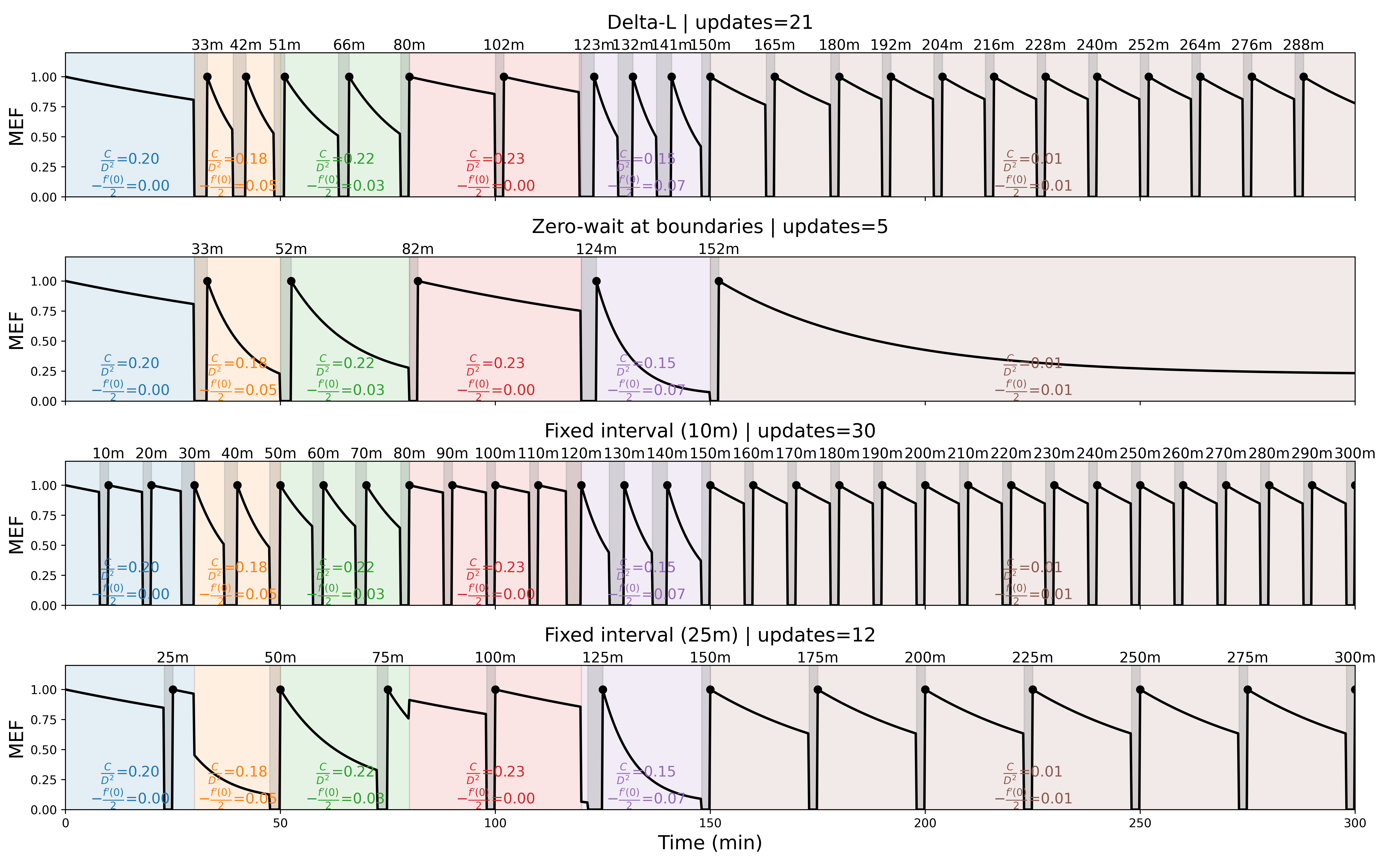}
    \caption{Comparison of Delta-L with baseline update policies. Delta-L optimally aligns updates with segment boundaries and fast-decay periods, achieving higher objective value $J$ than Zero-wait, Fixed-10m, and Fixed-25m baselines.}
    \label{fig:policies}
\end{figure*}
\par
Baseline strategies expose common pitfalls: Zero-wait over-conservatively maximizes work time but achieves the poorest efficacy, yielding a 33.9\% gap vs. Delta-P. Conversely, Fixed-10m over-aggressively incurs the highest cost  yet delivers average efficacy, demonstrating that update frequently without theory can't work. These results validate that environment-aware, cost-sensitive scheduling substantially outperforms the baselines.
\par
Fig. \ref{fig:policies} shows the specific update measures comparing the Delta-L schedule to three baselines. The optimal trajectory aligns several completions with boundaries where $T_j$ is large and $S_{\rm entry}$ is small, and introduces short delays when downtime $D_j$ is large or when a slight shift improves alignment with the next segment. The zero-wait baseline reacts only at boundaries and misses profitable within-segment refreshes; fixed-interval baselines ignore heterogeneity and often refresh during low-benefit intervals, which increases downtime without proportional efficacy gains. Across instances, Delta-L achieves higher $J$ than all baselines while keeping the number of updates moderate, with updates clustering in fast-decay segments and in front of strong entry losses.
\par 
To further understand the decision-making mechanism behind these results, we compare $\frac{C}{D^2}$ with $-\frac{f'(0)}{2}$ in Fig. \ref{fig:policies} and find that the short-run rule is a local, single-interval criterion at age 0. The actions of long-term strategies in some environmental segments does not conform to the theory of short-term strategies.
The short-run rule considers one interval in isolation and asks whether adding one more update inside this interval provides immediate benefits. It treats the current state as fresh and ignores how this update changes the starting state and the budget for later intervals. The long-run objective is global: it maximizes average benefit per unit cost over the whole horizon and applies the same benefit–cost yardstick to all updates. Because each update resets the state and shifts the timing of future intervals, the optimal plan moves updates toward periods with lower cost or higher marginal benefit, and may delay or skip updates that look attractive locally.
\par
Although the long-run optimal policy can depart from the short-run rule, the direction of choice is still well indicated by short-run intuition. Using this as a reference, we classify environments into the following types and, based on extensive simulations, report the update times most often chosen in each type.
\par
\noindent\textit{Type A (small-entry, fast such as in the train platform).}
With strong entry loss and fast intra-segment decay, ZERO\_WAIT appears with a \emph{nontrivial} probability, yet DELAYED still dominates in our default sampling. 
This is because the global optimum sometimes prefers a short delay to align with subsequent segments.

\noindent\textit{Type B (large-entry, slow such as walking in the street).}
DELAYED overwhelmingly dominates; ZERO\_WAIT is rare and NO\_UPDATE appears sparsely. 
This matches the intuition: weak entry loss ($S_{\rm entry}\approx 1$) and slow decay reduce the marginal benefit of immediate refreshes, while downtime and cost remain.

\noindent\textit{Type C (others).}
A mixed pattern emerges: when the draw falls into ``small-entry + medium/fast'' regimes, ZERO\_WAIT is more frequent; otherwise DELAYED dominates. 
NO\_UPDATE appears mainly in the ``large-entry + slow'' corner.
\par
\begin{figure}[!t]
    \centering
    \includegraphics[width=\columnwidth]{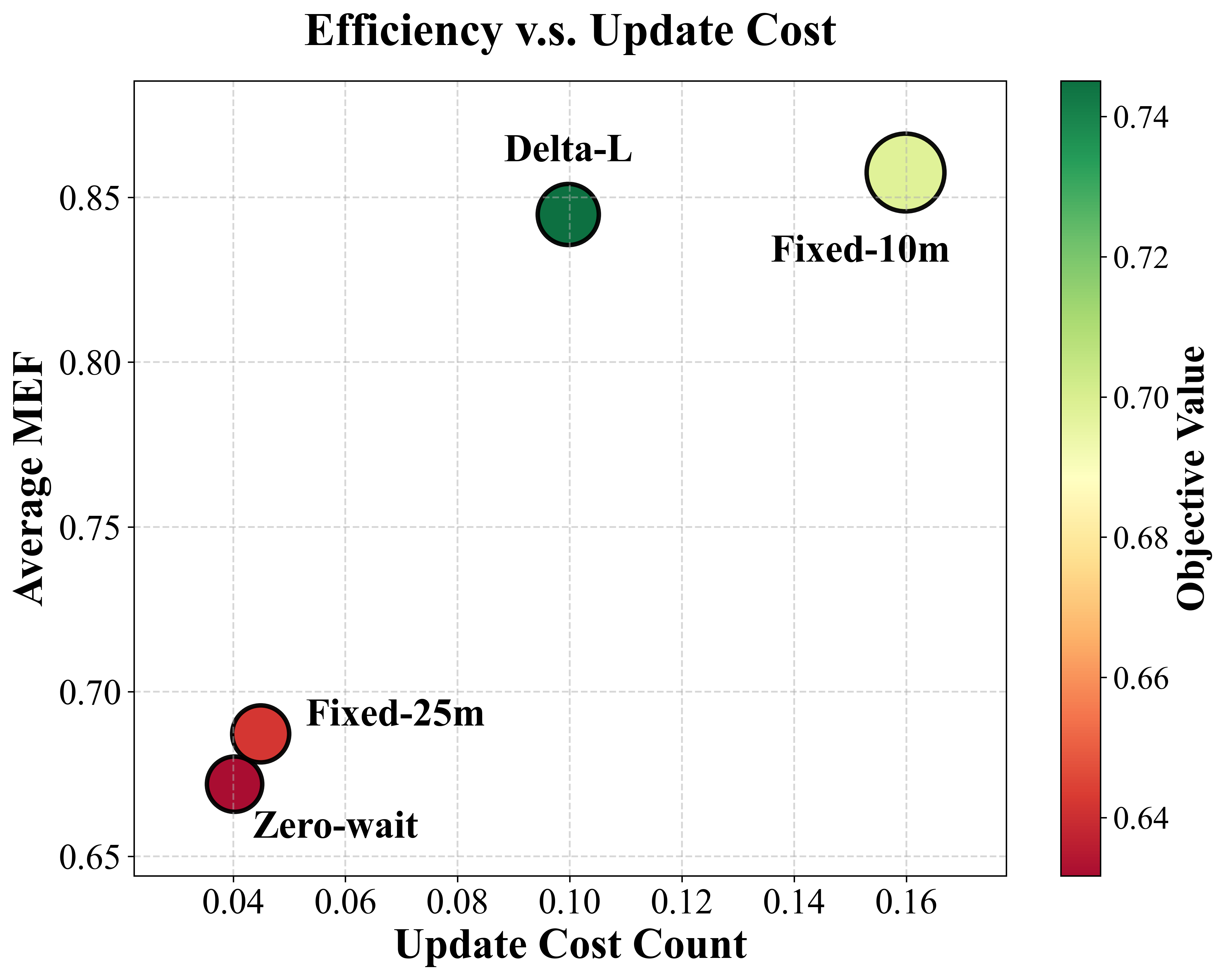} 
    \caption{Trade-off between average MEF and update cost, where the color of each point indicates its objective value.}
    \label{fig:performance_comparison}
\end{figure}
To evaluate the detailed trade-off between performance and cost, the average update cost, average MEF and update times are illustrated in Fig.~\ref{fig:performance_comparison}. While the aggressive Fixed-10m policy yields the highest MEF (0.86), it does so at the cost of highest update frequency. Delta-L, however, strikes a superior balance, achieving 98.6\% of the maximum MEF with 37.5\% fewer update cost and 33.3\% fewer update times. This highlights our algorithm's primary advantage: it intelligently allocates resources to attain near-optimal CKM quality while significantly reducing operational costs and system downtime, proving its efficacy for practical network deployments.

\section{Conclusion}
This paper addresses the critical question of \emph{when} to update a Channel Knowledge Map in dynamic environments. By introducing the Map Efficacy Function and formulating update scheduling as fractional programming, we develop Delta-P for global optimality and Delta-L for near-linear scalability.
Our results reveal that optimal update decisions are governed by the competition between information decay and resource consumption. For unpredictable environments, we derive a closed-form threshold: immediate updates are optimal when $-f'(0)/2 > C/D^2$. This reflects an intuitive trade-off—when information value decays faster than resources are consumed, maintaining freshness outweighs computational savings; otherwise, deliberate delay is preferable.
For predictable environments, stronger entry loss and faster decay favor immediate updates to prevent rapid performance degradation, while weaker entry loss and slower decay allow delayed updates to conserve resources. Notably, long-term strategies rationally deviate from short-term rules by aligning updates with segment boundaries and redistributing budget across the horizon, prioritizing global performance over local gains.

Future work includes multi-cell coordinated updating, online learning of environmental dynamics, and integration with resource allocation.

\bibliographystyle{IEEEtran}
\bibliography{ref}

\end{document}